# Input referred low-frequency noise analysis for single-layer graphene FETs

Nikolaos Mavredakis and David Jiménez

*Abstract*—The bias-dependence of input referred low-frequency noise (LFN), $S_{VG}$, is a considerable facet for RF circuit design. $S_{VG}$ was considered constant in silicon transistors but this was contradicted by recent experimental and theoretical studies. In this brief, the behaviour of $S_{VG}$ is investigated for single-layer graphene transistors based on a recently established physics-based complete compact LFN model. A minimum of $S_{VG}$ is recorded at the bias point where maximum transconductance is located which coincides with the peak of the well-known M-shape of the normalized output LFN and the model precisely captures this trend. Mobility fluctuation effect increases $S_{VG}$ towards to lower currents near charge neutrality point (CNP) while carrier number fluctuation and series resistance effects mostly contribute away from CNP; thus, $S_{VG}$ obtains a parabolic shape vs. gate voltage similarly to CMOS devices.

## I. INTRODUCTION

INPUT referred (or gate voltage) low-frequency noise (LFN) $S_{VG}$ is a crucial figure of merit (FoM) since it can provide advantageous information regarding the selection of the operating point of analog/RF circuits. It can be defined as [1]:

$$S_{VG} = \frac{S_{ID}^2}{g_m^2} \quad (1)$$

where $S_{ID}$ is the output or drain current LFN and $g_m$ is the transconductance of the device. Equation (1) has been proven to be valid in all regions of operation [2]. LFN model proposed in [2] is based on carrier number fluctuation (ΔN) effect which is generated by trapping/detrapping mechanism of active traps near the gate oxide of the device [3] and assume a $(g_m/I_D)^2$ bias-dependence of normalized output noise divided by squared drain current $(S_{ID}/I_D^2)$ where $I_D$ is the drain current of the transistor. Such approximation which is valid under the consideration of a uniform channel at low drain voltage $V_{DS}$ region, results in a constant $S_{VG}$ vs. gate voltage $V_{GS}$ (or $I_D$) because of (1), as shown in the following expression:

$$S_{VG} = \frac{S_{ID}^2}{g_m^2} = \frac{S_{ID}^2}{I_D^2}\frac{I_D^2}{g_m^2} = S_{VFB}\frac{g_m^2}{I_D^2}\frac{I_D^2}{g_m^2} = S_{VFB} \quad (2)$$

where $S_{VFB}$ is the flat-band voltage spectral density related to trapped charge density [2] and is a constant.

Two more mechanisms generate LFN to semiconductor devices; mobility fluctuation (Δμ) effect due to fluctuations of the bulk mobility which is described by the empirical Hooge formula [4] and contact resistance contribution (ΔR) [5]. Recent experimental findings from MOSFETs prove that the approximation of a constant $S_{VG}$ does not hold [5], [6]. An increasing trend of $S_{VG}$ is reported towards deeper weak inversion due to Δμ mechanism as well as a similar increase towards stronger inversion due to ΔN and ΔR effects which result in a minimum in moderate inversion. This can be a reliable indicator for circuit designers to select the specific regime to bias the circuits. As mentioned before, the simplified $(g_m/I_D)^2$ ΔN approach predicts a constant $S_{VG}$, thus the aforementioned increase due to ΔN effect in higher current regime of MOSFETs is attributed to Coulomb Scattering (CS) effect [2]. The same occurs for more complete ΔN models [5, §6.3.1], where the vast percentage of this increase is still induced by CS effect.

Nowadays, new material candidates are explored to replace silicon for integrated circuits (ICs) due to the limitations on gate length scaling in CMOS devices. Among them, graphene has been proven to be an ideal contestant for RF applications due to its remarkable characteristics [7]. Graphene transistors (GFETs) have been already fabricated and used in circuits at research labs [8] but for large scale integration production, LFN and more specifically $S_{VG}$ should be thoroughly investigated and this is the main goal of the present study. LFN is up-converted to phase noise and consequently, deteriorates the performance of analog/RF circuits such as oscillators [9] or terahertz detectors [10]. Several GFET LFN characterization and modeling studies have been reported [11]-[15] where simple $S_{ID}/I_D^2$ models [11], [12] based on ΔN $(g_m/I_D)^2$ approximation [2] have been proposed, whereas a complete physics-based compact model including all the LFN generators (ΔN, Δμ, ΔR) was developed by our group [14], [15] based on a chemical-potential IV model [16]. Our model is proven to be valid in all regions of operation as it includes the effect of non-homogeneous channel on LFN as well as the contribution of Velocity Saturation (VS) effect at high electric field regime. To our knowledge, no study has been reported yet regarding the behavior of $S_{VG}$ in GFETs. A constant $S_{input}$ is defined in [11] which is equivalent to $S_{VFB}$ mentioned before, but any possible bias-dependence of $S_{VG}$ has not been explored so far. In this work, we give insight into the performance of $S_{VG}$ at different operating conditions by validating our model with experimental data from three fabricated short-channel GFETs [17].

N. Mavredakis and D. Jiménez are with the Departament d'Enginyeria Electrònica, Escola d'Enginyeria, Universitat Autònoma de Barcelona, Bellaterra 08193, Spain. (e-mail: Nikolaos.mavredakis@uab.es).

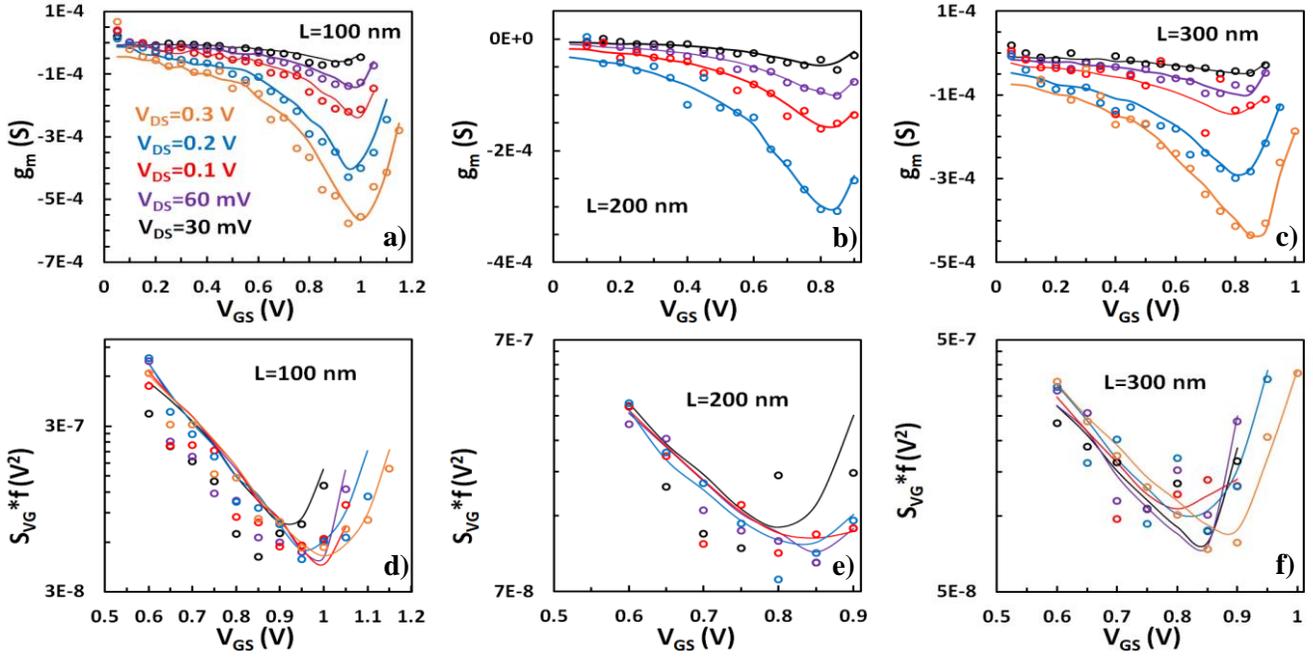

**Fig.1.** Transconductance $g_m$ (upper plots) and input noise $S_{VG}*f = S_{ID}*f/g_m^2$, referred to 1 Hz, (bottom plots) vs. gate voltage $V_{GS}$ with markers representing the measurements and lines the model for short-channel GFETs with W=12 μm and L=100 nm (a, d), L=200 nm (b, e) and L=300 nm (c, f), respectively for all available drain voltage $V_{DS}$ values.

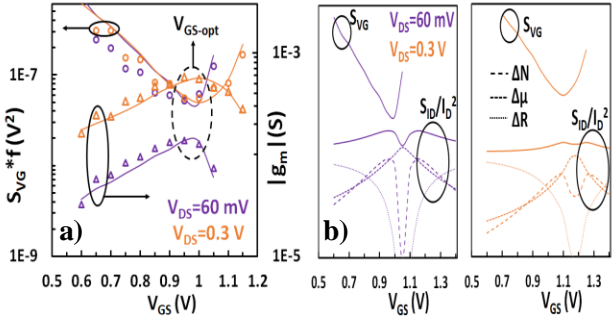

**Fig.2.** a) $S_{VG}*f$ (left y axis) and absolute $|g_m|$ (right y axis), b) $S_{VG}*f$, $S_{ID}*f/I_D^2$ (left y axis) vs. $V_{GS}$ for short-channel GFETs with W=12 μm and L=100 nm for $V_{DS}$=60 mV, 0.3 V. markers: measured, solid lines: total model, dashed and dotted lines: different noise contributors ΔN, Δμ and ΔR respectively.

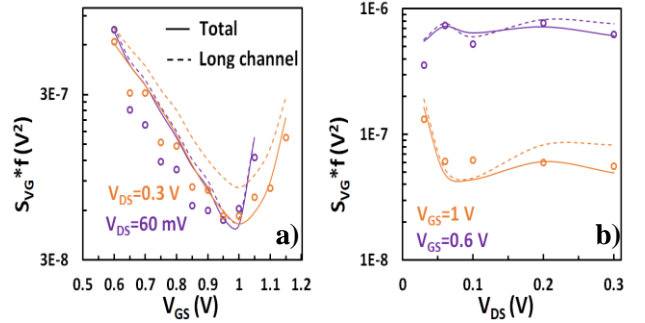

**Fig.3.** $S_{VG}*f$ vs. a) $V_{GS}$ for $V_{DS}$=60 mV, 0.3 V and b) $V_{DS}$ for $V_{GS}$=0.6 V, 1 V for short-channel GFETs with W=12 μm and L=100 nm. markers: measured, solid lines: total model, dashed lines: long-channel model.

## II. DEVICES AND MEASUREMENTS

$I_D$ and LFN measurements were conducted at three short-channel single-layer (SL) GFETs with width W=12 μm and gate length L=300 nm (A300), L=200 nm (A200) and L=100 nm (A100), respectively [14], [17]. The data were obtained after sweeping $V_{GS}$ from strong p-type to strong n-type region including charge neutrality point (CNP) for a wide range of $V_{DS}$ values, covering from low to high electric field regime. In the present work, p-type region is exclusively shown as maximum $g_m$ is reported there [14] while asymmetries are recorded in IV data between p- and n-type region [14] due to different electron and hole mobilities or due to effects caused by parasitic junctions at graphene metal contacts (For more details on the device fabrication process and schematics as well as the measurement setup, see [14], [17]). Regarding LFN, $S_{ID}$ was measured from 1.5 Hz to 2.5 KHz, averaged from 10-40 Hz in order to calculate its value at 1 Hz and then $S_{VG}$ data were estimated through (1) where $g_m$ is extracted as the first derivative of $I_D$ w.r.t. $V_{GS}$.

In order to calculate the simulated $S_{VG}$ through the $S_{ID}$ model [14], [15], the value of the modeled $g_m$ is needed in order to apply (1). Thus, it is crucial to have a precise fitting between measured and simulated $g_m$ and this is achieved successfully as it is illustrated in Fig. 1a (A100), 1b (A200) and 1c (A300), respectively where $g_m$ is plotted vs. $V_{GS}$ at all available $V_{DS}$ values. One parameter set is used for the IV simulations which can be found elsewhere [14].

## III. RESULTS – DISCUSSION

A complete bias-dependent analysis of experimental $S_{VG}$ data at f=1 Hz is presented in this section while the model is validated successfully at every operation regime. Measured $S_{VG}$ presents a minimum versus $V_{GS}$ for every device under test (DUT) at each individual $V_{DS}$ from low to high electric field resulting in a parabolic shape and the model precisely captures this trend (cf. Fig. 1d-1f). (extracted LFN model parameters can be found elsewhere [14].) It is also evident that the magnitude of this minimum $S_{VG}$ has a slight $V_{DS}$ dependence while the whole $S_{VG}$ shifts slightly to higher $V_{GS}$ as $V_{DS}$ increases which can be justified by the corresponding increment of $V_{CNP}$ [14]-[16]. In general, $S_{VG}$ is nearly independent of $V_{DS}$ towards p-type region below the $V_{GS}$ value where its minimum is recorded ($V_{GS}<V_{GS-MINSVG}$), and the model captures this trend. Looking thoroughly to the $V_{GS-MINSVG}$, one could notice that it is located in the same point where maximum $|g_m|$ is achieved ($V_{GS-MAXGM}=V_{GS-MINSVG}=V_{GS-opt}$) for all DUTs and this can be explained by (1) where $g_m$ is in the denominator. The latter is shown in Fig. 2a vs. $V_{GS}$ for the A100 GFET both low $V_{DS}$=60 mV and high $V_{DS}$=0.3 V. $S_{VG}$ is shown in the left y-axis while $|g_m|$ in the right. Up to now, $S_{ID}/I_D^2$ was shown to be maximum at the same preceding point where the peak of the well-known ΔN effect related M-shape is also reported [11]-[15] and this is confirmed by our simulations in Fig. 2b (left subplot: $V_{DS}$=60 mV, right subplot: $V_{DS}$=0.3 V, $S_{VG}$-$S_{ID}/I_D^2$: left y axis) vs. $V_{GS}$ where ΔN, Δμ, ΔR effects are also presented. The latter indicates that $V_{GS-opt}$ ensures minimum $S_{VG}$ appropriate for biasing circuits in terms of LFN performance despite the maximum $S_{ID}/I_D^2$ recorded there.

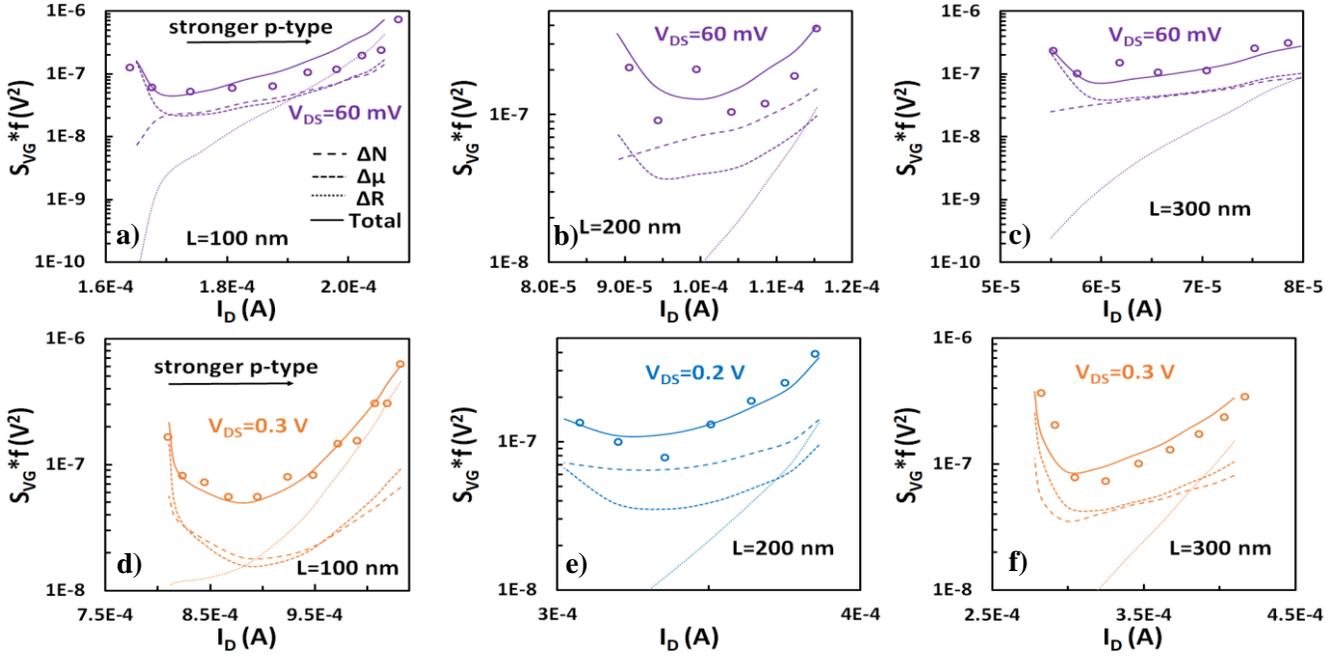

**Fig.4.** $S_{VG}*f$ vs. drain current $I_D$ for short-channel GFETs with W=12 μm and L=100 nm (a, d), L=200 nm (b, e) and L=300 nm (c, f), respectively for low $V_{DS}$=60 mV (upper plots) and higher $V_{DS}$=0.2, 0.3 V (bottom plots). markers: measured, solid lines: total model, dashed and dotted lines: different noise contributors ΔN, Δμ and ΔR respectively.

In Fig. 3 both measured and modeled $S_{VG}$ are presented vs. $V_{GS}$ (Fig.3a) and $V_{DS}$ (Fig.3b) for the A100 DUT whereas a long-channel model after de-activating VS effect [14], [15] is also depicted with dashed lines. Both low and high $V_{DS}$=60 mV, 0.3 V cases are drawn in Fig. 3a and a decrease of the total model (similarly to $S_{ID}/I_D^2$ [14], [15]) is observed at high $V_{DS}$ which is accurately confirmed by the experimental data. This reduction is associated with VS effect which has been successfully incorporated in our model [14], [15]. Oppositely, the long-channel model overestimates noise in the high $V_{DS}$ case while it coincides with the total model at low $V_{DS}$ where VS effect is insignificant. Regarding Fig. 3b, $S_{VG}$ is illustrated for two $V_{GS}$ values; near ($V_{GS}$=1 V) and away ($V_{GS}$=0.6 V) CNP. At higher $V_{DS}$, the complete model predicts successfully experimental $S_{VG}$ while the long-channel one overestimates it and the latter is more evident near CNP ($V_{GS}$=1 V). No strong deviations are observed to $S_{VG}$ with respect to $V_{DS}$ away from CNP confirming its independence of $V_{DS}$ while near CNP there is a steep decrease of $S_{VG}$ from $V_{DS}$=30 mV to 60 mV while it slightly changes for the rest of higher $V_{DS}$ values. This decrease is attributed to $V_{GS}$= 1 V being higher than $V_{GS-opt}$ at $V_{DS}$=30 mV and thus $S_{VG}$ has already started to increase in contrary with the rest of $V_{DS}$ cases which exhibit higher $V_{CNP}$ and consequently higher $V_{GS-opt}$ (cf. Fig. 1d).

In order to investigate deeper the contribution of each of ΔN, Δμ and ΔR mechanisms to $S_{VG}$, an explicit analysis is demonstrated in Fig. 4 where $S_{VG}$ is sketched vs. $I_D$ for low (cf. Fig. 4a-4c) and high (cd. Fig. 4d-4f) $V_{DS}$ for all the available GFETs. Apart from the measurements and the total model, each individual contribution is shown with dashed and dotted lines for every case. It is evident that ΔN effect increases $S_{VG}$ with $I_D$ away from CNP towards p-type region below $V_{GS-opt}$ for both low and high $V_{DS}$. It is worth mentioning that this increase is recorded without the impact of CS effect as in MOSFETs since the LFN model fits the data without the contribution of the aforementioned mechanism. Near CNP above $V_{GS-opt}$, ΔN contribution decreases as $I_D$ is lessened for low $V_{DS}$ which is reasonable as $S_{ID}/I_D^2$ is quite imperceptible there (deep minimum of M-shape, cf. Fig. 2b [14], [15]). In the contrary, at high $V_{DS}$, ΔN has a sizeable effect on $S_{VG}$ near CNP and adds to its increase with $1/I_D$ noticed there as the non-homogeneous channel results in an increase of the minimum of $S_{ID}/I_D^2$ M-shape, cf. Fig. 2b [14], [15] which apparently affects $S_{VG}$ also. ΔN contribution continues to grow with $I_D$, at stronger p-type region both at low and high electric fields where with the simultaneous effect of ΔR effect, a further and steeper boost is caused to $S_{VG}$ especially at high $V_{DS}$ region; $\Delta R$ is trivial near CNP. Regarding $\Delta\mu$ effect, it is mainly responsible for the increase of $S_{VG}$ with $1/I_D$ near CNP above $V_{GS\text{-}opt}$. $\Delta\mu$ main contribution to $S_{ID}/I_D^2$ has been shown to be near CNP [14], [15] (cf. Fig. 2b), and this is confirmed regarding $S_{VG}$ in the present study. $\Delta\mu$ effect is also noticed to increase with $I_D$ towards stronger p-type region below $V_{GS\text{-}opt}$ at both high and low $V_{DS}$ and thus, can moderately contribute to $S_{VG}$ even away CNP. The above theoretical observations are successfully verified on experimental data from all the DUTs (cf. Fig. 4.)

## IV. Conclusions

Input referred LFN is for the first time analyzed thoroughly for SL GFETs in this brief. The assumption of a constant $S_{VG}$ versus $V_{GS}$ is experimentally proven to be incorrect in GFETs, similarly to CMOS. Instead, a parabolic-like behavior is demonstrated around an optimum $V_{GS\text{-}opt}$ point where $|g_m|$ is maximized. The coexistence of maximum $|g_m|$ and minimum input LFN can certainly induce an attractiveness from the circuit design aspect regarding the specific point. We successfully validated our recently proposed LFN model with the above experimental findings where $\Delta\mu$ mechanism is proven to mainly increase $S_{VG}$ as $I_D$ is reduced near CNP above $V_{GS\text{-}opt}$ whereas below this, $\Delta N$ leads to an $S_{VG}$ increment towards stronger p-type region which becomes even steeper with the concurrent contribution of $\Delta R$ effect far away from CNP.

### Acknowledgements

We acknowledge Prof. Henri Happy, Associate Prof. Emiliano Pallecchi and Dr. Wei Wei from Carbon group (IEMN institute, University of Lille, France) for fabricating the GFET under test. This work was funded by the European Union's Horizon 2020 research and innovation program under Grant Agreement No. GrapheneCore2 785219 and No. GrapheneCore3 881603. It has also received partial funding from the Spanish Government under the project RTI2018-097876-B-C21 (MCIU/AEI/FEDER, UE); and partial funding from the ERDF allocated to the Programa Operatiu FEDER de Catalunya 2014-2020, with the support of the Secretaria d'Universitats i Recerca of the Departament d'Empresa i Coneixement of the Generalitat de Catalunya for emerging technology clusters to carry out valorization and transfer of research results. Reference of the GraphCAT project: 001-P-001702.

### References

[1] S. Christensson, I. Lundstrom and C. Svensson, "Low frequency noise in MOS transistors—I Theory," *Solid State Electr.*, vol. 11, no. 9, pp. 797-812, Sep. 1968, 10.1016/0038-1101(68)90100-7.
[2] G. Ghibaudo, O. Roux, Ch. Nguyen-Duc, F. Balestra and J. Brini, "Improved Analysis of Low Frequency Noise in Field-Effect MOS Transistors," *Physica Status Solidi (a)*, vol. 124, no. 2, pp. 571-581, April 1991, 10.1002/pssa.2211240225.
[3] A. L. McWhorter, "1/f noise and germanium surface properties," *Semiconductor Surface Physics*, pp. 207-228, 1957.
[4] F. N. Hooge, "1/f noise," *Physica B+C*, vol. 83, no. 1, pp. 14-23, May 1976, 10.1016/0378-4363(76)90089-9.
[5] C. Enz, and E. Vitoz, "Charge Based MOS Transistor Modeling," Chichester, U. K..:Wiley, 2006, 10.1002/047085546.
[6] D. M. Binkley, "Tradeoffs and Optimization in Analog CMOS Design," *John Wiley and Sons,* 2008, 10.1002/9780470033715.
[7] A. C. Ferrari *et al.*, "Science and technology roadmap for graphene, related two-dimensional crystals, and hybrid systems," *Nanoscale*, vol. 7, no. 11, pp. 4598-4810, Sep. 2014, 10.1039/C4NR01600A.
[8] F. Schwierz, "Graphene transistors," *Nature Nanotechnol.*, vol. 5, no. 7, pp. 487-496, Jul. 2010, 10.1038/nnano.2010.89.
[9] E. Guerriero, L. Polloni, M. Bianchi, A. Behnam, E. Carrion, L. G. Rizzi, E. Pop and R. Sordan, "Gigahertz Integrated Graphene Ring Oscillators," *ACS Nano*, vol. 7, no. 6, pp. 5588-5594, Jun. 2013, 10.1021/nn401933v.
[10] S. Castilla, B. Terres, M. Autore, L. Viti, J. Li, A. Y. Nikitin, I. Vangelidis, K. Watanabe, T. Taniguchi, E. Lidorikis, M. S. Vitiello, R. Hillenbrand, K. J. Tielrooij and F. H. L. Koppens, "Fast and Sensitive Terahertz Detection Using an Antenna-Integrated Graphene pn Junction," *Nano Letters*, vol. 19, no. 5, pp. 2765-2773, May 2019, 10.1021/acs.nanolett.8b04171.
[11] I. Heller, S. Chatoor, J. Mannik, M. A. G. Zevenbergen, J. B. Oostinga, A. F. Morpurgo, C. Dekker and S. G. Lemay, "Charge Noise in Graphene Transistors," *Nano Letters*, vol. 10, no. 5, pp. 1563-1567, May. 2010, 10.1021/nl903665g.
[12] A. N. Pal, S. Ghatak, V. Kochat, E. S. Sneha, A. Sampathkumar, S. Raghavan and A. Ghosh, "Microscopic Mechanism of 1/f Noise in Graphene: Role of Energy Band Dispersion," *ACS Nano*, vol. 5, no. 3, pp. 2075-2081, March 2011, 10.1021/nn103273n.
[13] A. A. Balandin, "Low-frequency 1/f noise in graphene devices," *Nature Nanotechnol.*, vol. 8, no. 8, pp. 549-555, Aug. 2013, 10.1038/nnano.2013.144.
[14] N. Mavredakis, W. Wei, E. Pallecchi, D. Vignaud, H. Happy, R. Garcia Cortadella, A. Bonaccini Calia, J. A. Garrido and D. Jimenez, "Velocity Saturation effect on Low Frequency Noise in short channel Single Layer Graphene FETs," *ACS Applied Electronic Materials*, vol. 1, no. 12, pp. 2626-2636, Dec. 2019, 10.1021/acsaelm.9b00604.
[15] N. Mavredakis, W. Wei, E. Pallecchi, D. Vignaud, H. Happy, R. Garcia Cortadella, N. Schaefer, A. Bonaccini Calia, J. A. Garrido and D. Jimenez, "Low-frequency noise parameter extraction method for single layer graphene FETs," *IEEE Trans. Electron Devices*, vol. 67, no. 5, pp. 93-99, Mar. 2020, 10.1109/TED.2020.2978215.
[16] D. Jimenez, and O. Moldovan, "Explicit Drain-Current Model of Graphene Field-Effect Transistors Targeting Analog and Radio-Frequency Applications," *IEEE Trans. Electron Devices*, vol. 58, no. 11, pp. 4377-4383, Nov. 2011, 10.1109/TED.2011.2163517.
[17] W. Wei, X. Zhou, G. Deokar, H. Kim, M. M. Belhaz, E. Galopin, E. Pallecchi, D. Vignaud and H. Happy, "Graphene FETs with Aluminum Bottom-Gate Electrodes and Its Natural Oxide as Dielectrics," *IEEE Trans. Electron Devices*, vol. 62, no. 9, pp. 2769-2773, Sept. 2015, 10.1109/TED.2015.2459657.